\documentclass[11pt]{article}
\usepackage{amsmath,amsthm,amsfonts,graphics,epsfig}
\usepackage{subfigure}
\usepackage{epsf}
\usepackage{graphicx}
\usepackage{graphics}
\usepackage{amsmath}
\usepackage{url}
\usepackage{epstopdf}
\newcommand{\N}{\mathcal{N}}

\newcommand{\A}{\mathcal{A}}
\newcommand{\D}{\mathcal{D}}

\newcommand{\norm}[1]{\left\Vert#1\right\Vert}
\newcommand{\be}{\begin{equation}}
\newcommand{\ee}{\end{equation}}

\newcommand{\abs}[1]{\left\vert#1\right\vert}
\newcommand{\Real}{\mathbb R}

\newtheorem{thm}{Theorem}[section]
\newtheorem{prop}{Proposition}[section]
\newtheorem{lem}{Lemma}[section]
\newtheorem{cor}{Corollary}[section]

\newtheorem{rem}{Remark}[section]

\def\mean{\hbox{${\rm I \hskip -2pt E}$}}
\def\rit{\hbox{\rm I\hskip -2pt R}}

\def\zit{\hbox{\rm Z\hskip -4pt Z}}

\def\notin{\hbox{$\in \hskip -10pt / \; $}}
\def\ind{\hbox{${\rm 1\hskip -3pt I}$}}
\def\mean{\hbox{${\rm I \hskip -2pt E}$}}
\def\Pr{\hbox{${\rm I \hskip -2pt P}$}}
\def\pr{\hbox{${\rm I \hskip -2pt P}$}}

\begin{document}

\begin{center}
{\Large {\bf Buffer occupancy asymptotics in rate proportional sharing networks with heterogeneous long-tailed inputs}}
\end{center}

\begin{center}
{ \bf Ozcan OZTURK\footnotemark[1], Ravi R. MAZUMDAR\footnotemark[2], Nikolay LIKHANOV\footnotemark[3]
}
\end{center}

\noindent \footnotemark[1] Qualcomm Research Center, Qualcomm Inc., San Diego, Ca. 92121, USA   (E-mail: ozcanoz@gmail.com)

\noindent \footnotemark[2] Electrical and Computer Engineering, University of Waterloo,
Waterloo, ON  N2L 3G1, Canada (E-mail: mazum@uwaterloo.ca)

\noindent \footnotemark[3] Institute for Problems of Information
Transmission, Russian Academy of Sciences, Moscow, Russia

\begin{center}
\today
\end{center}

\begin{abstract}
In this paper, we consider  a network of rate proportional processor sharing servers in which sessions with long-tailed duration
arrive as Poisson processes. In particular, we assume that a session of type $n$
transmits at a rate $r_n$ bits per unit time and lasts for a random time
$\tau_n$ with a generalized Pareto distribution given by $\pr \{\tau_n > x\} \sim
\alpha_n x^{-(1+\beta_n)}$ for large $x$, where $\alpha_n, \beta_n
> 0$. The weights are taken to be the rates of the flows. The network is assumed to be loop-free with respect to
source-destination routes. We characterize the order $O-$asymptotics of the complementary buffer
occupancy distribution at each node in terms of the input characteristics of the sessions.
In particular, we show that the distributions obey a power law whose exponent can be
calculated via solving a fixed point and deterministic knapsack problem. The paper
concludes with some canonical examples.
\vspace{0.3cm}

\emph{Keywords:} Queueing networks; Processor sharing; Buffer asymptotics; Pareto distribution; Long tailed distributions.
\end{abstract}

\section{Introduction}

Modern networks carry many different types of traffic streams with widely differing characteristics. The statistical 
variations that are of primary interest are those which are short lived and those that are long flows. The reason is that these two types of flows have very different impacts in terms of 
engineering the network in terms of router speeds and buffering requirements. The key issue is that long duration flows must be handled carefully and their statistical characteristics 
usually are the dominant ones in terms of network queueing performance. On the time-scale of long flows, short flows can be viewed as averaged out and while they can take up bandwidth 
it is predictable and thus the primary situation of interest is one in which long-lived flows are present.  A detailed statistical analysis of such sessions
suggests that they are well modelled by ON-OFF type of processes
where the ON periods,  of random duration, have a tail
distribution which decays according to a long-tailed distribution.
Different definitions are used in the literature for long and heavy tails.
Here we assume that a r.v. $X$ is long-tailed if
$\Pr \{ X> x \}\sim const. \ x^{-\beta}$ with $\beta > 0$.
If $\beta \in (0, 2]$, then $X$ has infinite variance and is said to heavy-tailed. These distributions 
are said to be sub-exponential.
When many independent such sessions arrive randomly, the
aggregated input process is long-range dependent.

The performance of networks and their
ability to offer Quality of Service (QoS) depend on accurately
capturing the parametric dependence of the QoS measures such as
the delay or loss distributions. While calculating these distributions exactly is
intractable, the asymptotics of the tail distribution are much more manageable.

Numerous studies have shown that the presence of long and heavy-tailed traffic
forces us to change the way the buffer dimensioning is done in that
much more buffering is necessary to achieve similar buffer
overflow characteristics than with conventional or short-range dependent traffic.
Indeed, the way buffer
overflows occur can be very different. This is essentially due to
the way that large excursions of the buffer workload take place.
For conventional traffic models, the tail of the stationary
workload distribution is exponential while for
long-range dependent traffic, it has an asymptotic long-tail or sub-exponential decay.

There are many results available for buffer asymptotics of a single node or server with
long-tailed and sub-exponential inputs. These have been obtained
under different hypotheses. These hypotheses relate to conditions
on the session length distribution decay rates, session
transmission rates and session arrival models. All results
deal with stationary queues where the average rate is less than
capacity. The vast majority of results are for FIFO systems where the
source transmission rates are identical, which is referred as the
homogeneous case. These can be found, for example, in
\cite{Duffield1,Jelenkovic,Nain,Heath,Makowski,Rolski}. An
excellent survey for the homogeneous case can be found in~\cite{Boxma}.  It has however been
shown in~\cite{LiMa} that assuming the same transmission rates can
lead to erroneous conclusions on the asymptotic behavior of the
tail distribution of the buffer length when the inputs have
long-tailed session lengths. This was obtained for the so-called
$M/G/\infty$ model where sessions arrive according to a Poisson
process, transmit at different rates and have differing long-tailed
random session holding times. Similar results and conclusions have also  been obtained
for a fixed number of ON-OFF sources in~\cite{BBJ}. Since these early papers there have been a number of works 
on studying the effects of sub-exponential distributions and how they arise in the context of network traffic, see \cite{Jelenkovic} and references therein.

The extension of above results to a  general network is quite
difficult since the traffic  loses its simple parametric structure
after passing through its entrance node. Networks with exponentially distributed service
times were considered in~\cite{Ananth96} and~\cite{Ganesh96}.
 In~\cite{Majewski96}, the author studied the large deviations problem for feedforward
 networks in heavy traffic.  The papers were basically concerned
with the computation of the rate functions associated with the tail distributions of the
buffer occupancy. This is a difficult variational problem. In
general,  when the input and output rate functions
 in queues have the so-called ``linear geodesics" an end-to-end analysis is feasible by an
iterative procedure. In~\cite{OconGan98,OCondep97}, it was shown
that in queues with many inputs the linear geodesic property does
not hold and thus calculating the rate functions of the outputs is difficult.

There are very few results  available for  networks with long-tailed service times.
In~\cite{Baccelli99}, the authors consider the (max, plus) setting and derive asymptotics of
the distributions of total response times for networks which include tandem queues.
In~\cite{Sigman99}, a feedforward network is analyzed
when a node with sub-exponential service time has upstream nodes with lighter-tailed service
times. In spite of these early successes the network case with general long-tailed have been analyzed only for very special cases such as feed-forward, or tandem networks \cite{Kim_09}. Most network models assume service times are i.i.d at nodes with sub-exponential distributions. In reality, flows do not change their sizes and thus the analysis must take into account the fact that within the networks the service times are no more independent.  A recent issue \cite{QUESTA_13} devoted to network asymptotics  does not address the issue of network buffer distributions. Thus, except for special network models that do not correspond to flow type session traffic there are no explicit results available that characterize the stationary buffer distributions.

In this paper we consider the model where sessions arrive independently as Poisson processes and transmit at a fixed
rate during their session duration. At a given node when the server capacity is exceeded, the capacity is shared in proportion to the rates of the flows else the source passes through unchanged. This is often referred to as rate proportional sharing and can be viewed as a discriminatory  processor sharing discipline (DPS) \cite{Altman_DPS} that approximates a weighted based round-robin service where the weights correspond to the rates of the flows entering the router. The rate multiplied by the number of flows denotes the aggregated required bandwidth of the flows.
The session or flow durations are assumed to be long-tailed,
having an asymptotic power law or
generalized Pareto decay. The sessions are assumed to be heterogeneous so that they have different
session duration distributions and different transmission rates. This gives rise to an heterogenous
 $M/G/\infty$
type of model for the inputs.  It is assumed that the types of sessions are defined by the routes that they take through the network and it is assumed that paths are loop-free. 
 Although the path for a single end-to-end route is loop free, the interaction between the flows in the network creates dependence between flows within the network, i.e., the input characterization is changed. In particular the probabilistic characterization  of the duration and rates changes due to the sharing of the processors.
We  obtain the
$O$-asymptotics  (i.e. we identify the asymptotic power law decay) for the buffer
occupancies at each node. In particular it shows that the characterization of the decay rate as the solution an optimal knapsack problem for a single server carries through to the network case with changes in the rates. The utility of this result is that we can identify the sources
that are the most problematic and also obtain rough estimates for the loss and delay
distributions. These results extend to networks with rate proportional sharing the results that were obtained in~ \cite{LiMa} however it is now more complicated due to the fact that output processes are difficult to characterize in non-Markovian settings. In the single node
case, exact asymptotics of buffer occupancy and loss
obtained in~\cite{LiMa} showed that these asymptotics are not
only governed by the tail distribution of sources but also
depend on their transmission rates that could be obtained from a knapsack problem. This result was also obtained in \cite{Zwart_2004} using the idea of equilibrium service times pioneered by Pakes \cite{Pakes} for tails sub-exponential inputs to a queue.

In this paper we show that a similar characterization  also holds
in a general loop-free network with rate proportional sharing. However computing the exponents of the buffer tail distributions need us to first solve a fixed point problem.
This is because the interaction of traffic
inside the network causes transmission rates and average load rates to be modified
during long buffer exceedance times. Furthermore, since we do not assume
that the network is feedforward, it results in a fixed point equation for computing  the ``modified" rates. It is difficult to improve these results (e.g. obtain
exact asymptotics)  since the Poisson arrival structure and
independence of sessions are lost for the traffic inside the network.

The organization of this paper is as follows: In Section
\ref{sectionModel}, we formulate the model and present the
preliminaries. Section \ref{sectionMainresults} contains the main
result with the proofs. An example of a two node network is
considered in Section \ref{sectionExamples}. In Section
\ref{sectionConc}, we give a discussion of results and concluding
remarks.

\section{Model and Preliminaries}
\label{sectionModel}

\begin{figure}[hbtp]
\begin{center}
\includegraphics[scale=0.5]{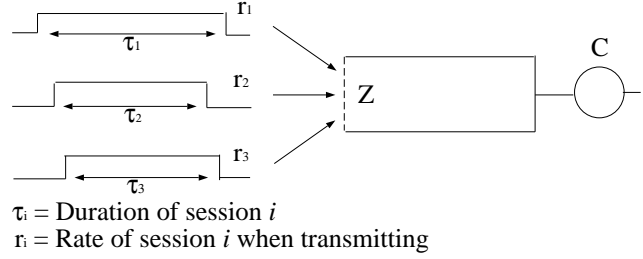}
\end{center}
\caption{Model of arriving sessions}
\label{Queue}
\end{figure}

We consider a discrete-time fluid DPS model where traffic
arrivals and services take place in slots indexed by $t \in \zit $
with the convention that arrivals take place at the beginning of a
slot and services are completed at the end of the slot.  This can be
seen as corresponding to a situation in the continuous time where the arriving traffic
coming in $(i,i+1]$ are all served at time $i+1$.
We refer
to $t$ as the time instant in the discrete-time model. There is a
finite set $\N$ of traffic types (classes) with $N = card(\N)$
which are differentiated according to their transmission rates and
their session lengths and different classes  are assumed to be
mutually independent. Session requests for type $n \in \N$ arrive
randomly according to a Poisson process with rate $\lambda_n$.
 Let $\theta ^n_t$ be the
number of sessions of class $n$ arriving at time $t$. We assume
that  $\theta ^n_t$ are i.i.d. for fixed $n$ and
\[
\pr \{ \theta ^n_t=k \} =e^{-\lambda _n} \frac{\lambda
_n^k}{k!}
\]
A session of class $n$ then transmits at the rate $r_n$ (bits/slot) for a duration $\tau_n$
(slots) which is assumed to have a long-tailed distribution. Let $\tau^n
_{t,j}$ denote the session length of the $j$'th session of class
$n$
 arriving at time $t.$ The r.v.'s $\tau^n_{t,j}$ are
assumed to be i.i.d. for fixed $n$ and satisfy
\[
\pr \{ \tau^n_{t,j} \geq z \} =\pr \{ \tau _{n}
\geq z \} \sim \alpha_n z^{-(1+\beta_n)}
\]
where $\alpha_n, \beta_n >0$ and $A(x)\sim B(x)$
 means that $\lim_{x\rightarrow \infty}
\frac{A(x)}{B(x)}=1$. Also $A(x) \preceq (\succeq) B(x)$ means
$\limsup_{x\rightarrow \infty} \frac{A(x)}{B(x)} \leq 1 (\liminf_{x\rightarrow \infty} \frac{A(x)}{B(x)} \geq 1)$,
i.e., the inequalities are in an
asymptotic sense.
This model is  a generalization of the $M/G/\infty$ model proposed
by Cox
\cite{Cox}, which we refer to as the $(M/G/\infty)^N$ model
and is depicted in Fig.~\ref{Queue}.

The network is composed of $M$ nodes (see Fig.~\ref{Pnetwork}
below). It is assumed that the packets from the sessions are admitted into an
infinite buffer and the buffer is served at a rate of $C_m$ per
unit time at node $m=1,\ldots,M$. The capacity is shared in proportion to the number of sessions and their rates. Type $n \in \N$ traffic has a
fixed route without any loops and its path is represented  by the
vector ${ \pi}^n =\left[\pi^n_1,\ldots, \pi^n_{l_n}\right]$ where
$\pi_i^n \in \{1,\ldots,M\}$. Hence, type $n$ traffic traverses the
nodes by entering the network at node $\pi^n_1$ and leaving  after
node $\pi^n_{l_n}$ and $\pi_i^n \neq \pi_j^n$ for $i \neq j$.
We define the set of traffic
types which  pass through node  $m$ by ${ \N_m} \doteq \{n \in \N: \ \pi_i^n =
m,  \ 1\leq i \leq l_n\}$.
\begin{figure}[h]
\begin{center}
\includegraphics[height=6.5cm, angle=0]{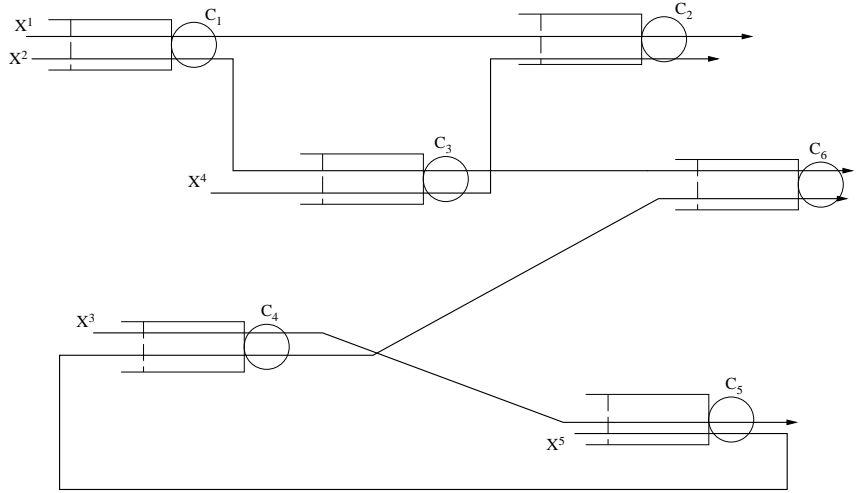}
\caption{\label{Pnetwork}A typical network considered in this paper}
\end{center}
\end{figure}

Let $X^n_t$ be the input traffic process of class $n$ entering the network. Then,
\[
X^n_t=\sum_{i=t}^{-\infty }\sum_{j=1}^{\theta^n _{i}}r_n{\bf 1}
_{\{\tau^n_{i,j}\geq t-i\}}.
\]
and average (mean) load of class $n$ is
\[
\rho_n \doteq \lambda_n r_n \mean[\tau_n]
\]
 
 Let $W^{m,n}_t$ denote the workload due to a flow of type $n$ at server $m$ and 
\[
W^m_t= \sum_{n\in \N_m} W^{m,n}_t
\]
 denote the total work due to all flows at server $m$. 
 
Let 
\begin{equation}
\label{stability}
\sum_{n \in \N_m} \rho_n < C_m;\ \forall \ m
\end{equation}

Under this hypothesis, the queues in the network are stable since the
network is loop-free for all classes of flows.
condition, by Loynes' theorem~\cite{Baccelli}, there exists a
stationary version of $W^m$.

Consider the following splitting of the input process $X^n_t$ into two processes
$X^{n,L}_t$ and $X^{n,H}_t$ as follows: The process $X^{n,L}_t
$  is formed by  active sessions which have
session lengths at most $\varepsilon z$ (will be referred as ``short" sessions
in the context) with $\varepsilon >0$ and  given by
\begin{equation}
\label{short} X^{n,L}_t =\sum_{i=t}^{t-\varepsilon z}
\sum_{j=1}^{\theta^n_i}r_n {\bf 1}_{\{\varepsilon z \geq \tau^n
_{i,j}\geq t-i\}}
\end{equation}

In the sequel, $z$ will be taken as the large buffer occupancy level and $\varepsilon >0$
will be chosen to be a small enough number. The process $X^{n,H}_t$ is composed of
sessions with  lengths greater than $\varepsilon z$ (will be referred as ``long" sessions in the sequel)
and given by
\begin{equation}
\label{long} X^{n,H}_t =\sum_{i=t}^{-\infty} \sum_{j=1}^{\theta^n_i}
r_n {\bf 1}_{\{\tau^n _{i,j} > \varepsilon z, \tau^n_{i,j} > t-i   \}}
\end{equation}

As it can be  seen from the definition, the processes
$X^{n,L}_t,X^{n,H}_t$ are mutually independent and $X^n_t =
X^{n,L}_t + X^{n,H}_t$. The superscript `$H$' indicates that the
process is strictly ``heavy" (or ``long") and the superscript `$L$' indicates that the
process is strictly ``light" (or ``short"). We will refer to $X^{n,L}$ and
$X^{n,H}$ as short and long processes in this context. As $z \rightarrow \infty ,$ we have
${\mean}X^{n,L}_t \rightarrow {\mean}X^n_t$
and ${\mean}X_t^{n,H}\rightarrow 0$. Despite the fact that ${\mean}
X^{n,H}_t \rightarrow 0$ as $z \rightarrow \infty$, the process
$X^{n,H}_t$ will contribute significantly to
the large buffer occupancy probability. 

Later, we will choose
$\varepsilon$ small enough so that the short sessions can be replaced with constant inputs
without any effect on the buffer length O-asymptotics.

We define $X^n_{k,t}$ ($Y^n_{k,t}$) to be the class $n$ input (output)
traffic at node $k$ and  time $t$. For any process $Z$ with value $Z_t$ at time $t$, the cumulative
process in the time interval $(t_1,t_2]$ will be denoted by
$Z(t_1, t_2)$, i.e.,
$
Z(t_1,t_2)=\sum_{i=t_1+1}^{t_2}Z_t.
$

We also need the following definitions to state the
main result. Let $J \in \zit_+^{N}$ where $\zit_+ = \{0,1,2,\ldots\}$. $J$ will
correspond to the combination of long (longer than $\varepsilon
z$) sessions; i.e. there are $J_n$ long sessions of class $n$. If
the transmission rates of active long sessions at a time exceed
 the difference of total output capacity and average load at a node, a
scaling effect would occur.
This happens because the service capacity is shared
among all these active sessions according to rate proportional scheduling and at times when
there is not enough output capacity at a node to serve all the traffic without buffering,
the output process of this node will behave as a scaled version of its input process.
In particular, the mean load of short sessions
(which goes to the average workload as $z\rightarrow \infty$) and
the transmission rates of long sessions would change at the
output.
In order to quantify this, we first define $p^n(m)=m'$ for  $n \in
\N_m$ if $\pi^n_i =m$ and $\pi^n_{i-1} = m'$; i.e. $p^n(m)$ is
the upstream node of node $m$ for class $n$. Take $p^n(m) = 0 $ if
$m =\pi^n_1 $. We define the transmission rate and mean of class
$n$ traffic entering node $m$ as follows:
 \be
 \label{ratescale}
r^{n,J}_m = \left\{ \begin{array}{ll}
0 & if \ n \not\in \N_m \\
r_n & if \ p^n(m)=0 \\
r^{n,J}_{p^n(m)} S^J_{p^n(m)} & otherwise
\end{array}\right.
 \ee
 and
 \be
 \label{rhoscale}
\rho^{n,J}_m = \left\{ \begin{array}{ll}
0 & if \ n \not\in \N_m \\
\rho_n & if \ p^n(m)=0 \\
\rho^{n,J}_{p^n(m)} S^J_{p^n(m)} & otherwise
\end{array}\right.
 \ee
where \be \label{scalefunc} S^J_m= \frac{C_m}{\max\{C_m, \sum_{n
\in \N_m } J_n r_m^{n,J} + \rho_m^{n,J} \}} \ee

$S^J_m$ is just the fraction of the capacity available to each flow at the node $m$ and so $r_nS^J_m$ would be the 
rate allocated to flow $n$ at node $m$ with configuration of $J$ long flows.
Note that the output
traffic is no longer $M/G/\infty$ . 

The function $S^J_m$ measures the
scaling of  sources for the output transmission at node $m$ for a combination of
$J$ active long sessions. Note that
$0< S^J_m  < 1 $  when the input rates
of the long sessions exceed the difference between the output capacity and
the total average loads at node $m$ and $S^J_m =1$ if this does not happen.
\begin{lem}
\label{lemmascale} 
For every $J \in \zit_+^{N}$, equations
(\ref{ratescale}) and (\ref{rhoscale}) are well defined, i.e.,
there exist unique solutions $r^{n,J}_m$ and $\rho^{n,J}_m$
satisfying these equations. Furthermore, for fixed $J$,
$r^{n,J}_m$ and $\rho^{n,J}_m$ are continuous in parameters $r_n$
and $\rho_n$.
\end{lem}
{\bf Proof.} See Appendix A.

\vspace{2mm}

We now define the following for future usage:
\begin{eqnarray}
\label{defnsJ1}
R^J = \sum_{n \in \N } J_n r_n, &
R^J_m = \sum_{n \in \N_m } J_n r_m^{n,J} \\
\label{defnsJ2}
\rho = \sum_{n \in \N } \rho_n, \vspace{5mm}&
\rho^J_m = \sum_{n \in \N_m } \rho_m^{n,J}\\
\label{defnsJ3}
\kappa_J = \sum_{n \in \N} J_n \beta_n, &
J^m_0 = \arg \min_J \{\kappa_J:  R^J_m > C_m - \rho_m^{n,J}  \} \\
\label{defnsJ4}
\beta_{min} = \min_{n \in \N}{\beta_n}, & r_{max} = \max_{n \in
\N}{r_n}
\end{eqnarray}

The results below, taken from~\cite{LiMa} (Lemma 3.3 and Lemma 3.4 therein),
provide bounds for the deviations of a short process
from its average rate.
\begin{lem}
\label{llower}

For any $n \in \N$  and   sufficiently small
$\delta _1>0,$ $\delta _2>0$ and sufficiently large $z$
\[
\pr \{  \inf_{t \geq 0} \{ X^{n,L}(-t,0) -(\rho_n -\delta
_1)t  \} <-\delta _2 z \} \leq e^{-O(z)}.
\]
\end{lem}

\begin{lem}
\label{lupper}

For any $n \in \N$ and  $\delta_2^0 > 0, \delta_1 >0, c_1> 0$ and \\ $0 <
\varepsilon < \min \{ \frac{\delta_2^0}{2(\rho_m + \delta_1)},
\min_i \frac{\beta_i^*}{2c_1r_i} \}$ where $\beta_i^* = \beta_i ,
0 < \beta_i <1$ and $\beta_i^* = \frac{1}{2}, \beta \geq 1$, if z
is  sufficiently large, then uniformly over all $\delta_2 >
\delta_2^0$

\[
\pr \{  \sup_{t \geq 0} \{ X^{n,L}(-t,0) > t(\rho_n +\delta
_1) \} > \delta_2 z \}
 \leq z^{1-c_1 \bar \delta_2}
\]
where $\bar \delta_2 \doteq \delta_2 - (\rho_m
+\delta_1)\varepsilon > 0$.

\end{lem}

\section{Main Result and Proofs}
\label{sectionMainresults}

Our goal is to obtain the $O$-asymptotics for the probability that
the buffer content at node $m$ exceeds a large value $z$. We will
use $f(z) = O(g(z))$ to mean $0 < \liminf_{z \rightarrow \infty}
f(z)/ g(z)  \leq \limsup_{z \rightarrow \infty} f(z)/ g(z)< \infty$ and
$f(z) = o(g(z))$ to mean $0 < \lim_{z \rightarrow \infty} f(z)/ g(z) = 0$.
It will be shown that the buffer of node $m$ will reach a large
value $z$ if there are enough number of long sessions which
contribute to this. During this time, the short sessions show average
behavior, i.e., produce traffic around their mean load level. Thus,
the short processes can be replaced with constant (equal to the mean load) processes in the
asymptotic regime without causing a significant change at large
occupancy levels of buffers.

We first review the single node result ($M=1$) from~\cite{LiMa} (Theorem 2.1 therein)
where exact buffer asymptotics were obtained.
A form of their result, sufficient for our purposes
in this paper, is
\be
\label{LiMaMain}
\pr \{ W^1_0 > z  \} = \pr \{ \sup_{t>0} \sum_{n\in \N_1} X^n(0,t) - C_1t > z \} =
O(z^{-\kappa_{J^1_0}})
\ee
It was shown in~\cite{LiMa} also that the most likely large buffer occupancy is generated by
a busy period when exactly $J^1_0$ long sessions are active during this busy period.
If there is only a single type of flow ($card(\N)=1$), then
$J^1_0 = \lceil (C_1 - \rho_1)/r_1 \rceil$. Hence
the buffer length decay rate  $\kappa_{J^1_0}$  depends not only on the decay rates
$\beta_n$'s but also on  the transmission rates $r_n$'s  unless $r_1> C_1 - \rho_1$
in which case $J^1_0=1$.

The basic idea of the proof is the following (that is valid at a single buffer with independent short and long processes ):

Let $\{Y_t\}$ denote the input process to a buffer that operates at rate $C_m$ and $Y^L$ and $Y^H$ refer to the short and long inputs defined in (\ref{short}) and (\ref{long}) and thus:
\be
Y_t= Y^L_t (\varepsilon)+ Y^H_t(\varepsilon)
\ee

with $\mean [Y_t]= \sum_{n\in \N_m} \rho_n < C_m$. In the following sketch of the ideas for simplicity  we set $\sum_{n\in \N_m} \rho_n= \rho$.

Having defined this decomposition of $\{Y_t\}$ the basic idea of the proof is as 
follows.

\vspace{0.4cm}
\subsubsection*{Upper bound}

For the upper bound, we choose $\varepsilon$ such that 
$\mean[Y^H_t(\varepsilon)] < C_m-\delta_1-\rho$ for some $\delta_1>0$ small. Such 
a choice always exists since $\mean[Y_t] = \rho < C_m$.

Now consider $\{Y^L_t(\varepsilon)\}$ as an input to a queue with service rate 
$\rho+\delta_1$ and $\{Y^L_t(\varepsilon)\}$ as an input to another queue with 
service rate $ C_m-\rho-\delta_1$. Then both queues are stable and let 
$\{W^{U,L}_t\}$ denote the stationary workload for the first queue and 
$\{W^{U,H}_t\}$ denote the stationary workload for the second queue.

Define $X(-t,0) = \sum_{j=-t}^0 Y_j$, and $X^i(-t,0) = \sum_{j=-t}^0 
Y^i_j(\varepsilon);\ i=L,H$.

Then the stationary workloads at time 0 are given by:
$$W_0 = \sup_{t \geq 0} \{X(-t,0) - Ct\}$$
$$W^{U,L}_0 = \sup_{t\geq 0} \{X^l(-t,0) - (\rho+\delta_1) t\}$$
and
$$W^{U,H}_0 = \sup_{t\geq 0}\{X^h(-t,0) - (C-\rho-\delta_1) t\}$$
and hence it readily follows that
$$W_0 \leq W^{U,L}_0 + W^{U,H}_0$$ 
Therefore:
\begin{equation}
\pr(W_0 > z) \leq \pr(W^{U,L}_0 + W^{U,H}_0 > z)
\end{equation}

We can show that the complementary distributions  
 are such that: $\pr(W^{U,L}_0 > z) \sim o\left(\pr(W^{U,H}_0 > 
z)\right)$ and $\pr(W^{U,H}_0 > z)$ is long-tailed.
Noting that $W^{U,L}_0$ and $W^{U,H}_0$ are independent we can then use the 
following result which can be found in Feller \cite[p. 271, Vol. 2]{feller} 
which we state in a slightly extended form below:

\begin{prop}
\label{fellerprop}
Let $F_1(.)$ and $F_2(.)$ be two distribution functions such that as $x \to 
\infty$
$$ 1 - F_i(x) \sim a_i x^{-\nu_i}$$

Then the convolution $G = F_1*F_2$ satisfies, for $x \to \infty$
$$1 - G(x) \sim \alpha x^{-min(\nu_1,\nu_2)}$$
and $\alpha$ corresponds to the $\alpha_i$ associated to the index of the 
exponent.

Moreover, as $x \to \infty$
$$1-G(x) \sim (1-F_1(x)) + (1- F_2(x))$$
\end{prop}

Applying the above result it readily follows that:
$$\pr(W_0 > z) \leq \pr(W^{U,H}_0 >z ) (1 + o(z))$$

One of the key results that was established in \cite{LiMa} was the notion of an {\em Isolated Typical Busy Period} which is a busy period associated with the arrival of exactly $J_0$ long sessions where $J_0$ is defined as the minimum exponent given by:

\begin{description}
\item{$\kappa_J$} = $\sum_{i=1}^N \beta_i j_i$ which corresponds to decay 
exponent associated with the set $J$.

\item{$R_J$} = $\sum_{i=1}^N r_i j_i$ the rate corresponding to the set $J$ of 
sessions.

\item{$J_0$} = $(j_1^o,\ldots , j_N^o ) = argmin_J \{\kappa_J\ : R_J-(C-\rho) > 
0\}$

\end{description}

The idea of an ITBP is similar to the role of the{\em Principle of the Single Large Jump} in the context of sums of sub-exponential random variables as presented in the monograph by Foss {\em et. al} \cite{Foss}.
\vspace{0.4cm}

\subsubsection*{Lower bound}

The proof of the lower bound requires a finer argument. Choose $\delta_1> 0$ and 
then $\varepsilon> 0$ such that : $\mean[Y^L_t(\varepsilon)] > \rho - \delta_1$ 
and $\mean[Y^H_t(\varepsilon)] < C_m-\rho + \delta_1$. Once again for any 
$\delta_1 > 0$ there exists $\varepsilon > 0$ such that these conditions can be 
met since $\mean[Y_t] = \rho$. Now consider  two queues with inputs 
$\{Y^L_t(\varepsilon)\}$, $\{Y^H_t(\varepsilon)\}$ and service rates 
$\rho-\delta_1$ and $C-\rho+\delta_1$ respectively. Note that the queue 
corresponding to  first input is unstable.

Now with similar notation as above:
$$\sup_{t\geq 0}\{X(-t,0) -Ct\} \geq \sup_{t\geq 0}\{X^H(-t,0) - (C-\rho 
+\delta_1)t\} + \inf_{t\geq 0} \{X^L(-t,0) - (\rho -\delta_1)t\}$$

Denote by $W^{L,H}_0$ the stationary workload given by:
$$W^{L,H}_0 = \sup_{t\geq 0} \{X^h(-t,0) -(C-\rho+\delta_1)t\}$$
and let
 $$W^{L,L}_0 = \sup_{t\geq 0} \{ (\rho -\delta_1)t - X^l(-t,0)\}$$

Note, $W^{L,L}_0$ can be viewed as a risk process \cite{Asmussen} where 
$X^L(-t,0)$ corresponds to the cumulative claims in $[-t,0]$. Under the assumed 
condition we know that the ruin time is finite w.p.1.

From above:
$$\pr(W_0 > z) \geq \pr(W^{L,H}_0+W^{L,L}_0 > z)$$

Once again since $W^{L,L}_0$ and $W^{L,H}_0$ are independent r.v.'s, by 
establishing the fact that 
$\pr(W^{L,L}_0 > z) \sim o\left( \pr(W^{L,H}_0 > z)\right)$  by virtue of Proposition \ref{fellerprop} it 
follows that as $z\to \infty$ 
$$\pr(W^{L,H}_0 > z)(1+ o(z)) \leq \pr(W_0 > z)$$
establishing a way to obtain the lowerbound. By choice of $\delta_1\to 0$ we 
will then establish that the upper and lower bounds are equal within factors of 
$o(z)$.

\vspace{0.5cm}

Having given the basic idea of the proof we now proceed with the proofs of the 
principal results. In the network case we will show that
a short process of type $n$ as it enters the network
 can be replaced with a constant process $\rho_n +\delta$ ($\rho_n -\delta$)
 for small enough $\delta > 0$ if $n \in \N_m$ ($n \not \in \N_m$) while considering the asymptotics
 of node $m$.
 
 Then we will show that the busy period where the buffer of node $m$ reaches a large $z$ is created
by a combination of long sessions ($J$) which satisfies $R^J_m > C_m -
\rho_m^{n,J} $. Note that the scaling defined before Lemma
\ref{lemmascale} will take place at a node if long sessions have a
total rate more than the difference of the output capacity and the average
input load at this node. This scaling causes a long session to become
longer at the output with a smaller transmission rate. Also its transmission will start
after all the traffic in the buffer is served. Therefore
we will consider a process which is obtained from the long process
by shifting a session by the buffer occupancy at its arrival time
and lengthening it by an amount proportional to the total
traffic arriving during the lifetime of this session.
 This shifting and scaling will allow us to bound the time
when the service of a session finishes at the downstream nodes.
It will be shown that the probability that this new process has
$J$ active sessions at $t=0$ is $O(z^{-\kappa_J})$. Furthermore,
the case $R^J_m > C_m - \rho^J_m$ will dominate the other
combinations in probability when the buffer occupancy at node $m$
is greater than a large $z$. The optimal (asymptotically most likely) such
combination was defined to be as  $J^m_0$ with a
probability of $O(z^{-\kappa_{J^m_0}})$ and this will give the
upper bound.

The lower bound part is relatively easier. We assume that the
$J^m_0$ long sessions arrive in $(-kz, k(1-\alpha)z)$ where $0<
\alpha<1$ is small and $k>0$ is later chosen large enough. This event has a
probability of $O(z^{-\kappa_{J^m_0}})$. From the upper bound
part, the probability that a buffer has a level greater than
$\delta_1 z $ at time ($-kz -1$) is $z^{-\gamma}$ for some $\gamma
> 0$. For  small enough  $\delta_1$, the contribution of buffer
contents at time ($-kz -1$) to the buffer content of node $m$ at
$t=0$ will be less than $\delta_2 z$ for small $\delta_2>0$ with high probability.
The short sessions are again
replaced with constant processes but this time with $\rho_n -
\delta$ if $n \in \N_m$ and $\rho_n + \delta$ if $n \notin \N_m$.
Then we show that this combination would produce enough amount of traffic to
make the  buffer at node $m$ bigger than $z$.

Now we state the main result of the paper:

\begin{thm}
 (Buffer asymptotics in networks with fixed routes)
Consider a rate proportional server sharing network with fixed routes and independent $(M/G/\infty)^N$ heterogeneous external inputs with fixed routes specified by with rate parameters $r_i$, power law tails for the holding times specified by  $\beta_i+1$ at the start node of route $i$.

Let $\N_m$ denote the flows that use server $m$ and let  $r_m^{n,J}$ and $\rho_m^{n,J}$ denote the scaled rates and average loads specified by the fixed points in Lemma \ref{lemmascale}.

Define:
\begin{eqnarray}
R^J_m = \sum_{n \in \N_m } J_n r_m^{n,J} &
\rho^J_m = \sum_{n \in \N_m } \rho_m^{n,J}& \kappa_{J_m} = \sum_{n \in \N_m} J_n \beta_n \\
&J^m_0 = \arg \min_J \{\kappa_{J_m}:  R^J_m > C_m - \rho_m^{J}&  \} 
\end{eqnarray}

and assume $J^m_0$ is unique.

Let $W^m_0$ be the stationary workload of node $m$.
 Then as $z \rightarrow \infty$,
\[
\pr \{W^m_0 > z\} = O(z^{-\kappa_{J^m_0}})
\]
\end{thm}

{\bf Proof.}

{\it \underline{Upper bound:}}
\vspace{2mm}

The main idea is to show that if the buffer occupancy  at time $t=0$
for node $m$ reaches a large value of  $z$, then there must have
been enough number of long sessions contributing to this.
Consider a long session of type $n$ at time $t$ with length
$\tau^n_{t,j}> \varepsilon z$ for $j = 1, \ldots,\theta^n_t$.
Define $S^{n,0}_{t,j} = t$ and if node $m_2$ is downstream node of node
$m_1$ for type $n$ flow, define $S^{n,m_2}_{t,j}=  S^{n,m_1}_{t,j} +
W^{m_2}_{S^{n,m_1}_{t,j}}/ C_{m_2}$. If $m_2$ has no predecessor,
set $m_1 =0$. Note that
 a long session of type $n$ arriving to the network at time $t$ starts being served at node
$m$
 at time $S^{n,m}_{t,j}$.
Consider a fictional queue with infinite buffer and service
capacity $\varepsilon_1>0$ which serves the long sessions arriving
at node $m$. Let $\bar S^{n,m}_{t,j}$ be the time when the service
of the  considered long session ends at this fictional queue. Also
define
\[
\xi^{m}_{n} = \sum_{t=-1}^{-\infty} \sum_{j= 1}^{\theta^n_t}
\ind_{A_{t,j}^{n,m}}, \ \ A_{t,j}^{n,m} = \{ S^{n,m}_{t,j} < 0,
\bar S^{n,m}_{t,j} > 0, \tau^n_{t,j} > \varepsilon z \}
\]
Let $\xi^m = (\xi^m_n)$. We claim that \be \label{ximJ}
\pr \{\xi^m = J \} = O(z^{-\kappa_J}) \ee Then we will write
\[
\pr\{W_0^m > z\} \leq \pr \{ R^{\xi^m}_m > C_m - \rho^{\xi^m}_m \}+
\pr \{ W_0^m > z, R^{\xi^m}_m < C_m - \rho^{\xi^m}_m \}
\]
For the first term of right side,
\[
\pr \{ R^{\xi^m}_m > C_m - \rho^{\xi^m}_m \} = \sum_{J \in \zit_+^N}
\pr \{\xi^m = J, R^J_m > C_m - \rho^J_m \} \leq
\pr \{\xi^m = J \}
\]
From equation (\ref{ximJ}) and definition of $J^m_0$, we get
\[
\pr \{ R^{\xi^m}_m > C_m - \rho^{\xi^m}_m \} = O(z^{-
\kappa_{J^m_0}})
\]
We will then show that \be \label{WzRless} \pr \{ W_0^m > z,
R^{\xi^m}_m < C_m - \rho^{\xi^m}_m\} = o(z^{- \kappa_{J^m_0}}) \ee
and this will complete the proof of the upper bound.

Let us prove claim (\ref{ximJ}) now. Define $X_t \doteq \sum_{n \in \N}
X^n_t $. First,
for any $d>0$, there exists
$K>0$ such that \be \label{XtbigKt} \pr \{ X(t, t+\tau) > K\tau \}
\leq o(\tau^{-d}) \ee as $\tau \rightarrow \infty$. Indeed, consider a
queue with service rate $K-1 > \rho$ with input $X_t$ and let $W'_t$ be its stationary buffer
content. Then by
using (\ref{LiMaMain}), we get
\[
\pr \{X(t, t+\tau) > K\tau \} = \pr \{ X(-\tau,0) > K\tau \} \leq
\pr \{W'_0 > \tau \} =O(\tau^{-\kappa_{J_0}})
\]
where $J_0= \arg \min_J \{\kappa_J:  R^J > K-1 - \rho  \}.$
Define
$\abs{J} = \sum_{n \in \N} J_n $. But $\kappa_{J_0} / \beta_{min} > \abs{J_0}$ and
$\abs{J_0} > \frac{K-1-\rho}{r_{max}}$.
Thus we can choose $K$
such that $\kappa_{J_0} > d$, proving (\ref{XtbigKt}).

Now define
\[
\tilde A_{t,j}^{n,m} = \{ S^{n,m}_{t,j} < 0, \bar S^{n,m}_{t,j} >
0, \tau^n_{t,j} > \varepsilon z, \bar S^{n,m}_{t,j} -
S^{n,\pi^n(1)}_{t,j} < (cK)^M(\tau^n_{t,j})\}
\]
where $c>1/C_k$ for all $k=1,\ldots, M$ and $c > 1/\varepsilon_1$. Here $cK$ will be the
lengthening factor due to the scaling that a long session
will experience while going through a node, where $K$ is determined by the amount of traffic
arriving during its transmission.
  Also define $ \tilde \xi^{m}_{n} =
\sum_{t=-1}^{-\infty} \sum_{j= 1}^{\theta^n_t} \ind_{\tilde
A_{t,j}^{n,m}}, $ By using (\ref{XtbigKt})and choosing $K$ large enough, we conclude
\be
\label{tildeeqv}
\pr \{ \xi^m = J \} \sim \pr \{ \tilde \xi^m = J \}
\ee
thus proving equation (\ref{ximJ}). Now let $D = (cK)^M +1$. Define $W_t=(W^1_t, \ldots, W^M_t)$.
Then,
\[
\pr \{ \tilde \xi^m = J \} =
\sum_{ x \in \Real_+^{M\abs{J}} }
\sum_{- {\bf t} \in Z_+^{ \abs{J} } }
\pr  \{ W_{t_i} = x_i, c\abs{x_i} + D \tau^{n_i}_{ {\bf t}_i } > -{\bf t}_i,
\tau^{n_i}_{{\bf t}_i} > \varepsilon z\}
\]

\[
\hspace{5.5cm} i=1,\ldots, \abs{J}, n_i \in \N,
\sum_{i} \ind_{\{n_i =n\}} = J_n )
\]
Above, $x =(x_1, \ldots, x_{\abs{J}})$ where $x_i \in \Real_+^M$.
Now note that $W_{t_i}$ and $\tau^{n_i}_{ t_i }$ are independent
since $W_{t_i}$ is determined by arrivals before $t_i$. It is easy
to see that
\[
\sum_{- t \in Z_+ } \pr \{x + D \tau^{n}_{ t } > -t, \tau^{n}_{t} >
\varepsilon z\}= O( z^{-\beta_n})
\]
Furthermore since $W_t$ is stationary, $\pr \{W_{t_i} = x_i \} =
\pr \{ W_{\tilde t_i} = x_i \}$ where \\ $\tilde t_i = t_i - \min_i t_i$.
Therefore,
\[
\pr \{ \tilde \xi^m = J \} = \sum_{ x \in \Real_+^{M\abs{J}} }
\sum_{- {\bf t} \in Z_+^{ \abs{J} } } \pr \{W_{\tilde t_i} = x_i \}
O(z^{-\kappa_{J}}) = O(z^{-\kappa_{J}})
\]

Now we will prove equality (\ref{WzRless}).
Consider the system where all the buffers are empty at time $-T$ and
let $W_{T,t}=(W^1_{T,t},\ldots, W^M_{T,t})$ be the buffer occupancy (workload)
at time $t$ for this system. It is known that $W_{T,t} \rightarrow W_t$ a.s.
as $T \rightarrow \infty$.
Therefore we will first show that equality (\ref{WzRless}) holds when
$W^m_0$ is replaced by $W^m_{T,0}$ and then take the limit $T \rightarrow \infty$.

 Let us investigate the total arrival
traffic to node $m$ in the time interval $(-T,0)$ assuming that all the queues are
empty at $-T$ and
$R^{\xi^m}_m < C_m - \rho^{\xi^m}_m$. Choose $\varepsilon_1 > 0$
such that $2\varepsilon_1 > C_m - \max \{R^J_m + \rho^J_m \ : \
R^J_m + \rho^J_m < C_m \}$. If a long session is not in
$A_{t,j}^{n,m}$, then it had been served before $t=0$ in the
fictional queue defined above. Consider another queue with
capacity $C_m - \varepsilon_1$ which serves all of the remaining traffic.
Note that the buffer content of this queue at
$t=0$ is bigger than $W_{T,0}^m$ provided that $R^{\xi^m}_m < C_m -
\rho^{\xi^m}_m$.

Now we assume that the long sessions in
$A_{t,j}^{n,m}$ are active in $(-T,0)$ all the time and
the remaining long sessions of other classes are removed. Note that this does not decrease the
buffer content at node $m$. More generally, replacing flows accessing node $m$ by pathwise
larger ones and other flows by smaller ones does not decrease the buffer content at node $m$.
This can be seen by a sample path argument. If all the queues are empty at $t=-1$, then
from the arguments in Lemma \ref{lemmascale}, the claim will hold at $t=0$ and an induction on
$t$ will complete the proof.
Under the above assumption, let
$\bar W^m_{T,0}$ be the buffer content of the queue with capacity $C_m
- \varepsilon_1$. Then,
\[
\pr \{ W_{T,0}^m > z, R^{\xi^m}_m < C_m - \rho^{\xi^m}_m \} \leq
\pr \{ \bar W_{T,0}^m > z, R^{\xi^m}_m < C_m - \rho^{\xi^m}_m \}
\]
Define $\D_m = \{R^{\xi^m}_m < C_m - \rho^{\xi^m}_m,
\xi^m \mbox{ sessions are active in } (-T,0) \} $.
Also let $X_{m} \doteq  \sum_{n \in \N_m} X_{m}^n $ be
the total cumulative input to node $m$.
Now assume that
\be
\label{up}
\pr \{ \sup_{0\leq t  \leq T}
X^n_k(-T,-t) - (r^{n,\xi^m}_k + \rho^{n,\xi^m}_k
+\varepsilon)(T-t) > \varepsilon_2 z, \D_m  \} \preceq z^{-d}
\ee
and
\be
\label{down}
\pr \{ \inf_{0 \leq t \leq T}
X^n_k(-T,-t) - (r^{n,\xi^m}_k + \rho^{n,\xi^m}_k  -
\varepsilon)(T-t) < -\varepsilon_2 z, \D_m  \} \preceq z^{-d}
\ee
for any $d>0$. Note that this is true for $k = \pi^n_1$.
Indeed, if $\D_m$ holds, then the long processes are constant $(=
r^{n,\xi^m}_k)$ in $(-T,0)$ for type $n$ flow. Furthermore, from
Lemmas~\ref{llower} and~\ref{lupper}, a short process $X^{n,L}$
differs from the constant processes $\rho_n-\delta$ and $\rho_n +
\delta$ with probability $o(z^{-\bar d})$ for any $\delta> 0$.
Here $\bar d$ can be chosen arbitrarily big.
Then,
 \[
\pr \{ \sup_{0\leq t_1  \leq T} Y^n_k(-T,t) -
( \frac{r^{n,\xi^m}_m + \rho^{n,\xi^m}_m}{\max(C_k,R^{\xi^m}_m +
\rho^{\xi^m})}
 + \varepsilon_1 )(T-t) >
 \varepsilon_2 z,
\D_m   \}
\]
\[
\leq
\pr \{ \sup_{0 \leq t  \leq T} X^n_k(-T,-t) -
(r^{n,\xi^m}_k + \rho^{n,\xi^m}_k +\varepsilon_3)(T-t) >
\varepsilon_4 z, \D_m  \}+
\]
\[
\sum_{j \in \N_k, j \neq n}
\pr \{ \inf_{0 \leq t  \leq T}
X^j_k(-T,-t) - (r^{j,\xi^m}_k + \rho^{j, \xi^m}_k -
\varepsilon_3)(T-t) - \varepsilon_4 z, \D_m  \}
\]
\[\leq o(z^{-d})
\]
We will choose $\varepsilon_3 = \varepsilon_1/card(\N_k)$ and
$\varepsilon_4 = (\varepsilon_2- M \delta_1) /card(\N_k)$. Above
we used the fact that the output of node $k$ in the interval
$(-T, -t]$ is equal to the  arrival in an interval $(-T,
t']$. Note that the time being discrete is not a problem here
since there is an $\varepsilon_2 z$ term and $z$ is taken to be
large. The same argument can be used to show that
\[
\pr \{ \inf_{0\leq t \leq T} Y^n_k(-T,-t) -
(\frac{r^{n,\xi^m}_m + \rho^{n,\xi^m}_m}{\max(C_k,R^{\xi^m}_m +
\rho^{\xi^m}_m)}
  - \varepsilon_1 )(T-t) < -
 \varepsilon_2 z,
\D_m  \}
\]
\[
\leq o(z^{-d})
\]
From Lemma~\ref{lemmascale}, the equations~(\ref{up}) and~(\ref{down}) should hold
for all $n,k$.
This can be more formally shown by considering the values
\[
\bar \alpha(n,k) \doteq
\]
\[
\hspace{-0.7cm}\lim_{z \rightarrow \infty} \frac{1}{\log z} \log  \left \{\pr \{ \sup_{0\leq
t  \leq T}
X^n_k(-T,-t) - (r^{n,\xi^m}_k + \rho^{n,\xi^m}_k
+\varepsilon_1)(T-t) > \varepsilon_2 z, \D_m  \} \right\}
\]
and defining similarly $\underline{\alpha}(n,k)$ where $\sup$ is replaced with $\inf$.
These expressions evaluate to $\infty$ when $k = \pi^n_1$. Above arguments  show that
it is also true for all $n,k$ at the outputs assuming it for the inputs. Thus
$\bar \alpha(n,k) =\underline{\alpha}(n,k) =\infty$ is a fixed point solution of these
relations as given in Lemma
\ref{lemmascale}.
Since there can
only be one solution, we conclude that the equation~(\ref{up}) is valid for all $n,k$.
Thus for the input to node $m$, we have
\[
\begin{array}{l}
\pr \{ \sup_{0<t \leq T} X_m(-t,0) - (R^{\xi^m}_m + \rho^{\xi^m}_m +
\varepsilon_1)t > \varepsilon_2 z, \D_m  \} \\\
\leq \pr \{ X_m(-T,0) - (R^{\xi^m}_m + \rho^{\xi^m}_m +
\varepsilon_1)T > 0.5 \varepsilon_2 z, \D_m  \}\\
\ \ \ + \ \pr \{ \inf_{0<t \leq T} X_m(-T,t) - (R^{\xi^m}_m + \rho^{\xi^m}_m +
\varepsilon_1)(T-t) <- 0.5 \varepsilon_2 z, \D_m  \}\\
\preceq o(z^{-d})
\end{array}
\]
for any $d> 0$.
Remember that $\rho^{n,J^m_0}_m$ and
$r^{n,J^m_0}_m$ are continuous functions of $\rho_k, \ r_k$'s.
Then,
\[
\pr \{ \bar W_{T,0}^m > z, R^{\xi^m}_m < C_m - \rho^{\xi^m}_m, T_0 < T\}
\]
\[
\leq
\pr \{ \sup_{0<t \leq T} X_m(-t,0) - (R^{\xi^m}_m + \rho^{\xi^m}_m +
\varepsilon_1)t > \varepsilon_2 z, \D_m  \} \preceq o(z^{-d})
\]
and finally by taking $T \rightarrow \infty$, we get
\[
\pr \{ W_0^m > z, R^{\xi^m}_m < C_m - \rho^{\xi^m}_m \} \leq
o(z^{-\kappa_{J^m_0}})
\]
and this proves the claim in equation (\ref{WzRless}).

\vspace{2mm}

{\it \underline{Lower bound:}}
\vspace{2mm}

Let $A^J(x,\alpha)$ be the event that exactly $J$ long sessions
start in $(-x, -(1-\alpha)x]$ and are still active at $t=0$ where $0 <
\alpha < 1$. First note that, for every class $n$, the number of
such sessions is a Poisson r.v. with parameter
$\sum_{t=-x+1}^{-(1-\alpha)x} \lambda_n \pr \{\tau_n > -t \} $ which is $O(x^{-\beta_n})$.
From the independence of classes, it is easy to see that $\pr \{A^{J}(x,\alpha) \}
\sim O(x^{-\kappa_{J}})$.
We will now show that
\[
\pr \{A^{J^m_0}(x,\alpha) \} \preceq \pr \{W^m_0 > z\}
\]
for some $\alpha>0$ and $x=bz$ with $b>0$. First, from the upper
bound proof,
\begin{equation}
\label{bufferx} \pr \{W^k_{-x} > \delta_1 z\} \preceq O(z^{-\gamma})
\end{equation}
for some $\gamma > 0$ and any $k = 1, \ldots, M$. Note that $\{W^k_{-x} >\delta_1 z \}$ and
$A^{J^m_0}(x,\alpha)$ are independent because $W^k_{-x}$ is
determined by sessions which arrived before $-x+1$. Now define the
following two events:
\[
\begin{array}{l}
B_1 = \{ \mbox{there are active long sessions at time }t=-x  \}
\\
B_2 = \{ \mbox{a long session other than } J_0^m \mbox{ arrives
between } t=-x \mbox{ and } t=0  \}
\end{array}
\]
The arrival process of long sessions is also Poisson with rate
$\lambda_m \pr \{\tau_m > \epsilon z \}$. Since the arrivals at
different times are independent, it is easy to see that
\[
\pr \{B_1 | A^{J^m_0}(x,\alpha) \} \leq O(z^{-\beta_{min}}) \ \mbox{ and } \
\pr \{B_2 | A^{J^m_0}(x,\alpha) \} \leq O(xz^{ -\beta_{min}  - 1} ).
\]
These and equation~(\ref{bufferx}) give \be \label{eqnAjWB1B2} \pr
\{ A^{J^m_0}(x,\alpha) \} \sim \pr \{A^{J^m_0}(x,\alpha), \bar B_1,
\bar B_2, W^k_{-x} < \delta_1 z, \forall k=1,\ldots,M \} \ee where
$\bar B_1, \bar B_2$ are complements of $B_1, B_2$. In other
words, we can assume that all the buffers are   almost empty at
$t=-x$ regarding their contribution to a buffer level exceeding $z$ at
$t=0$  and the only active long sessions during $(-x,0)$ are the
ones of $J_0^m$. Let
\[ \A_m =
\{A^{J^m_0}(x,\alpha), \bar B_1, \bar B_2, W^k_{-x} < \delta_1 z,
\forall k=1,\ldots,M  \}
\]
Let $T = (1-\alpha)x$. Now assume that
\[
\hspace{-0.5cm}
\pr \{ \sup_{0\leq t_1
\leq t_2 \leq T} X^n_k(-t_2,-t_1) - (r^{n,J^m_0}_k +
\rho^{n,J^m_0}_k +\varepsilon)(t_2-t_1) > \varepsilon_2 z, \A_m  \}
\preceq z^{-d}
\]
and
\[
\hspace{-0.5cm}
\pr \{ \inf_{0\leq t_1 \leq t_2 \leq T}
X^n_k(-t_2,-t_1) - (r^{n,J^m_0}_k + \rho^{n,J^m_0}_k  -
\varepsilon)(t_2-t_1) < -\varepsilon_2 z, \A_m  \} \preceq z^{-d}
\]
for any $d>0, \varepsilon, \varepsilon_2 >0$. Note that this is true when $k = \pi^n_1$.
Indeed, if $\A_m$ holds, then the long processes are constant $(=
r^{n,J^m_0}_k)$ in $(-T,0)$ for type $n$ flow. Furthermore, from
Lemmas~\ref{llower} and~\ref{lupper}, a short process $X^{n,L}$
differs from the constant processes $\rho_n-\delta$ and $\rho_n +
\delta$ with probability $o(z^{-\bar d})$ for any $\delta> 0$.
Here $\bar d$ can be chosen arbitrarily big and since $T =
(1-\alpha)bz$, we can take $d < \bar d -1$. Then,
\[
\hspace{-0.5cm}
\pr \{ \sup_{0\leq t_1 \leq t_2 \leq T} Y^n_k(-t_2,-t_1) -
(\frac{r^{n,J^m_0}_m + \rho^{n,J^m_0}_m}{\max(C_k, R^{J^m_0}_m +\rho^{J^m_0}_m)}
 + \varepsilon_1 )(t_2-t_1) >
 \varepsilon_2 z,
\A_m  \}
\]
\[
\leq
\pr \{ \sup_{0\leq t_1 \leq t_2 \leq T} X^n_k(-t_2,-t_1) -
(r^{n,J^m_0}_k + \rho^{n,J^m_0}_k +\varepsilon_3)(t_2-t_1) >
\varepsilon_4 z, \A_m  \} +
\]
\[
\hspace{-0.4cm}\sum_{j \in \N_k, j \neq n} \pr \{ \inf_{0\leq t_1 \leq t_2 \leq T}
X^j_k(-t_2,-t_1) - (r^{j,J^m_0}_k + \rho^{j, J^m_0}_k -
\varepsilon_3)(t_2-t_1) - \varepsilon_4 z, \A_m  \}
\]
\[ \leq o(z^{-d})
\]
We will choose $\varepsilon_3 = \varepsilon_1/card(\N_k)$ and
$\varepsilon_4 = (\varepsilon_2- M \delta_1) /card(\N_k)$. Above,
we used the fact that the output of node $k$ in the interval
$(-t_2, -t_1]$ is equal to the  arrival in an interval $(-t_2',
t_1']$. Note that the error induced here due to the discreteness of
time is not a problem
since there is an $\varepsilon_2 z$ term and $z$ is taken to be
large. The same argument can be used to show that
\[
\hspace{-0.7cm}
\pr \{ \inf_{0\leq t_1 \leq t_2 \leq T} Y^n_k(-t_2,-t_1) -
(\frac{r^{n,J^m_0}_m + \rho^{n,J^m_0}_m}{\max(C_k + R^{J^m_0}_m +
\rho^{J^m_0}_m)}
  - \varepsilon_1 )(t_2-t_1) < -
 \varepsilon_2 z,
\A_m  \}
\]
\[
\leq o(z^{-d})
\]
Then by using Lemma~\ref{lemmascale} as was done in
the upper bound part, we conclude
\[
\hspace{-0.5cm}
\begin{array}{l}
\pr \{ X_m(-T,0) - (R^{J^m_0}_m + \rho^{J^m_0}_m -\varepsilon_1)T <
-\varepsilon_2 z- \alpha T (R^J + \rho^J + 1), \A_m  \} \\
\leq o(z^{-d})
\end{array}
\]
for any $d> 0$.
Then
\[
\begin{array}{l}
\hspace{-0.5cm} \pr \{W_0^m > z \} \geq \pr \{X_m(-x,0) > C_m x + z   \}
\geq \pr \{ X_m(-x,0) > C_m x + z,  \A_m \}
\\
\succeq \pr \{ (R^{J^m_0}_m + \rho^{J^m_0}_m -\varepsilon_1 -
\delta_3)T > C_m x +(1+ \varepsilon_2)z, \A_m \} + o(z^{-d})
\\
 \succeq \pr \{A^{J^m_0}(x,\alpha)\}
\end{array}
\]
For the last inequality, we used  $R^{J^m_0}_m + \rho^{J^m_0}_m >
C_m$ and chose $\alpha, \varepsilon_1, \varepsilon_2$ small enough
and $b$ large enough so that
\[
(R^{J^m_0}_m + \rho^{J^m_0}_m)(1-\alpha) > C_m + \frac{1-
\varepsilon_2}{b}
\]
Combining above with equation (\ref{eqnAjWB1B2}) gives
\[
\pr \{W_0^m > z\} \succeq \pr \{A^{J^m_0}(x,\alpha)\} \sim
O(z^{-\kappa_{J^m_0}})
\]
\qed

\begin{rem}
The uniqueness of $J^m_0$ was assumed for convenience and is not necessary. It can easily be seen that
the results and proofs still hold when there are many optimal configurations  of long sessions.
\end{rem}
\begin{rem}

If there are input flows to the network with light tailed session
lengths such that $\lim_{x \rightarrow \infty} \log \pr \{\tau > x\}/ \log x = -\infty $, then
they can be ignored in calculating the buffer asymptotics. Such an
input can be replaced with a Pareto tailed version with
arbitrarily large value $\beta$ without decreasing the probability
of large buffer occupancies. Therefore, in determining $J^m_0$, none of
these inputs need to be considered.
\end{rem}
\begin{rem}
If the session  lengths are regularly varying
(i.e. $\lim_{x \rightarrow \infty} \pr \{\tau_n > tx\} / \pr \{\tau_n > x\} = t^{-(1+\beta_n)}$
for $\beta_n> 0$), it can be shown that the buffer occupancy
distributions are also regularly varying with the parameters as found in the main result.
\end{rem}

We can also find the joint distribution of buffer asymptotics. We
only state the result since the proof follows mutatis mutandis as
above.
\begin{cor}
Let $\mathcal{S} \subset \{1,\ldots, M\}$ and $a_m > 0, m \in
\mathcal{S}$. Then as $z \rightarrow \infty$,
\[
\pr \{W^m_0 > a_m z, m \in S\} \sim O(z^{-\kappa_{\mathcal{S}}})
\]
where $\kappa_{\mathcal{S}} = \max \{\kappa_{J^m_0} \ | \ m \in
\mathcal{S}\}$
\end{cor}

\section{Examples}
\label{sectionExamples}

In order to illustrate the results, in particular how the rates
$r^{n,J}_m $, $\rho^{n,J}_m$  are determined, we consider two
examples of a simple network with two nodes. The first example is
a feedforward network while the second example is a network where
individual routes have no loops but the network is not
feedforward.

{\it Example 1:} There are three classes of traffic, one of which
uses resources from both nodes. The schema is illustrated in the
figure below.

\begin{figure}[h]
\begin{center}
\includegraphics[height=4cm, angle=0]{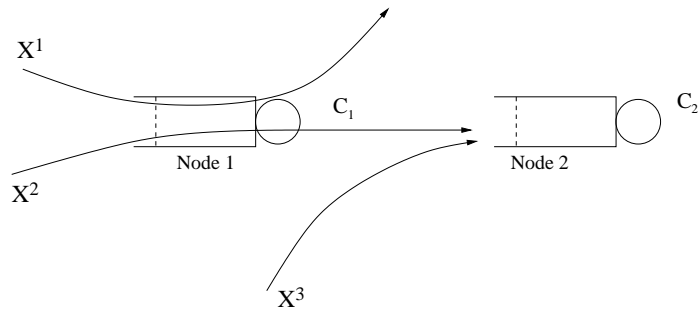}
\caption{Two node feedforward network}
\end{center}
\end{figure}

Now we calculate the large buffer asymptotics at the second node.
First,
\[
\begin{array}{lll}
r^{1,J}_2 = 0, & r^{2,J}_2 = \frac{r_2 C_1}{\max (J_1 r_1 + J_2
r_2 +\rho_1+ \rho_2,C_1)},
& r^{3,J}_2 = r_3 \\
\rho^{1,J}_2 = 0, & \rho^{2,J}_2 = \frac{\rho_2 C_1}{\max (J_1 r_1
+ J_2 r_2 + \rho_1+ \rho_2,C_1)}, & \rho^{3,J}_2 = \rho_3
\end{array}
\]
and thus
\[
\hspace{-0.5cm}
J^2_0=  \arg \min_{J \in Z_+^3} \{ \sum_{m=1}^3 J_i \beta_i  \ : \
\frac{(J_2 r_2+ \rho_2) C_1}{\max (C_1, J_1 r_1 + J_2 r_2 +\rho_1
+\rho_2)} +J_3 r_3 +\rho_3 > C_2
 \}
\]
In special cases, $J^2_0$ can be obtained easily. For example, if classes $2$ and $3$
have the same transmission and decay rates ($r_2 = r_3, \beta_2 = \beta_3$), then one
optimal configuration is
$J^2_0 = (0,0, \lceil \frac{C_2 - \rho_2 - \rho_3}{ r_3}\rceil)$.
This is because class $1$  sessions do not contribute to the buffer occupancy at node $2$ and
we can take all the long sessions to be of class $3$.

{\it Example 2:} In this example, we consider a non-feedforward
network with two nodes and two types of traffic as illustrated in
Fig. 4.
\begin{figure}[h]
\label{fignonfeed}
\begin{center}
\includegraphics[height=2cm, angle=0]{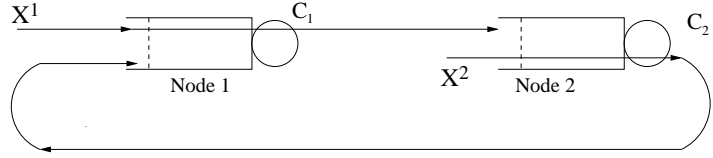}
\caption{Two node non-feedforward network}
\end{center}
\end{figure}

In this case, we obtain the following:
\[
\begin{array}{ll}
r^{1,J}_2 = r_1/ {\max (J_1 r_1 + J_2 r^{2,J}_1 +\rho_1+
\rho^{2,J}_1,C_1)}, & r^{2,J}_2 = r_2,
\\
r^{2,J}_1 = r_2/ {\max (J_1 r^{1,J}_2 + J_2 r_2 + \rho_2+
\rho^{1,J}_2,C_2)}, & r^{1,J}_1 = r_1,
\\
\rho^{1,J}_2 = \rho_1/ {\max (J_1 r_1 + J_2 r^{2,J}_1 +\rho_1+
\rho^{2,J}_1,C_1)}, & \rho^{2,J}_2 = \rho_2,
\\
\rho^{2,J}_1 = \rho_2/ {\max (J_1 r^{1,J}_2 + J_2 r_2 + \rho_2+
\rho^{1,J}_2,C_2)}, & \rho^{1,J}_1 = \rho_1,
\end{array}
\]
Hence \be \label{minkappaJ2} \kappa_{J^2_0}=  \min \{ \sum_{m=1}^3
J_i \beta_i  \ : J \in \Omega_1 \cup \Omega_2 \} \ee where
\[
\begin{array}{l}
\hspace{-0.7cm}
\Omega_1 = \{ J \in \zit_+^3 : J_1 r_1 + \rho_1+
\frac{(J_2r_2+\rho_2)C_2}{J_1r_1 +\rho_1 + J_2 r_2 +\rho_2} < C_1,
J_1r_1 +\rho_1 + J_2 r_2 +\rho_2 > C_2 \}
\\
\\
\hspace{-0.8cm}
\Omega_2 = \{ J \in \zit_+^3  :  J_1 r_1 + \rho_1+  \beta
(J_2r_2+\rho_2) > C_1, \ \alpha (J_1 r_1 + \rho_1) + J_2r_2 +
\rho_2 > C_2 \}
\end{array}
\]
Here $\alpha = C_1/(J_1 r_1 + \rho_1 + \beta (J_2r_2 + \rho_2)  )$
and $\beta = C_2/(\alpha(J_1 r_1 + \rho_1) + J_2r_2 + \rho_2)$.
Note that $\Omega_1$  corresponds to the situation where large buffer
occupancy happens only at the second node and $\Omega_2$ is where it happens
at both nodes. In finding the optimal point in equation
(\ref{minkappaJ2}), we need to check whether a combination $(J_1,
J_2)$ is inside $\Omega_1$ or $\Omega_2$. For $\Omega_1$, this is
easy. For $\Omega_2$, first $\alpha$ and $\beta$ are algebraically
or numerically solved and then substituted to the constraint in
the definition of $\Omega_2$.
If classes $1$ and $2$
have the same transmission and decay rates ($r_1 = r_2, \beta_1 = \beta_2$), then
it can be seen that it is enough to carry
out the above calculations assuming only the long sessions of class $2$.
In addition, if $C_1 > C_2$, then $J^2_0 = (0, \lceil \frac{C_2 - \rho_1 - \rho_2}{ r_2}\rceil)$
since now the most likely large buffer occupancy at the second node will happen  without one at the first node.

\section{Conclusion}
\label{sectionConc} In this paper we  considered a loop-free
networks of rate proportional servers in discrete-time with heterogeneous $(M/G/\infty)^N$
inputs with long-tailed Pareto holding times. We have shown that the
buffer occupancy distribution is also Pareto-like and determined its parameters. The
continuous time case can be readily treated with slight changes in
the technical details of the proofs. On the other hand, it seems
difficult to obtain finer asymptotics, i.e., determining the
prefactor constants because we cannot parametrically model the
traffic inside the network.

A further extension of the
model to cope with more general models will require different
tools since our proofs  depended  very heavily on the compound
Poisson structure of the input processes. Nevertheless the Poisson arrival of flows or sessions is by well motivated by the Poisson superposition 
theorem for stationary independent point processes.

\bibliography{all}

\begin{thebibliography}{10}

\bibitem{QUESTA_13}
Queueing systems (questa): Special issue on stochastic networks: Tail
  asymptotics for stationary distributions and related topics.
\newblock volume Volume 74, Issue 2-3, pages 105--368. 2013.

\bibitem{Altman_DPS}
E.~Altman, K.~Avrachenkov, and U.~Ayesta.
\newblock A survey on discriminatory processor sharing.
\newblock {\em Queueing Systems}, 53(1-2):53--63, 2006.

\bibitem{Ananth96}
V.~Anantharam.
\newblock Networks of queues with long-range dependent traffic streams.
\newblock In {\em Stochastic networks}, volume 117 of {\em Lecture Notes in
  Statist.}, pages 237--256. Springer, New York, 1996.

\bibitem{Asmussen}
S.~Asmussen.
\newblock {\em Applied probability and queues}, volume~51 of {\em Applications
  of Mathematics (New York)}.
\newblock Springer-Verlag, New York, second edition, 2003.

\bibitem{Baccelli}
F.~Baccelli and P.~Br\'emaud.
\newblock {\em {Elements of Queueing Theory, Series in Applications of
  Mathematics 26}}.
\newblock Springer-Verlag, N.Y., 1994.

\bibitem{Baccelli99}
F.~Baccelli, S.~Schlegel, and V.~Schmidt.
\newblock {Asymptotics of stochastic networks with subexponential service
  times}.
\newblock {\em Queueing Systems}, 33, No. 1-3:205--232, 1999.

\bibitem{BBJ}
S.~Borst, O.~Boxma, and P.~Jelenkovic.
\newblock {Generalized processor sharing with long-tailed traffic sources}.
\newblock In P.~Key and D.~Smith, editors, {\em Teletraffic Engineering in a
  Competitive World}, pages 345--354. Elsevier, 1999.

\bibitem{Boxma}
O.~J. Boxma and V.~Dumas.
\newblock {Fluid queues with long-tailed activity period distributions}.
\newblock {\em Computer Communications}, 21:1509--1529, 1998.

\bibitem{Cox}
D.~R. Cox.
\newblock {Long-range dependence: A review}.
\newblock In H.~A. David and H.~T. David, editors, {\em Statistics: An
  appraisal}. Iowa State University Press, Ames, Iowa, 1984.

\bibitem{Deimling}
K.~Deimling.
\newblock {\em Nonlinear functional analysis}.
\newblock Springer-Verlag, Berlin, 1985.

\bibitem{Duffield1}
N.~G. Duffield and N.~O'Connell.
\newblock Large deviations and overflow probabilities for the general
  single-server queue, with applications.
\newblock {\em Math. Proc. Cambridge Philos. Soc.}, 118(2):363--374, 1995.

\bibitem{feller}
W.~Feller.
\newblock {\em An introduction to probability theory and its applications, Vol.
  II}.
\newblock Wiley, N.Y., 1966.

\bibitem{Foss}
S.~Foss, D.~, Korshunov, and S.~Zachary.
\newblock {\em An introduction to heavy-tailed and subexponential
  distributions}.
\newblock Springer Series in Operations Research and Financial Engineering.
  Springer, New York, 2011.

\bibitem{Ganesh96}
A.~Ganesh and V.~Anantharam.
\newblock {Stationary tail probabilities in exponential server tandems with
  renewal arrivals}.
\newblock {\em Queueing Systems}, 22:203--248, 1996.

\bibitem{OconGan98}
A.~J. Ganesh and N.~O'Connell.
\newblock The linear geodesic property is not generally preserved by a {FIFO}
  queue.
\newblock {\em Annals of Applied Probability}, 8(1):98--111, 1998.

\bibitem{Gantmacher}
F.~R. Gantmacher.
\newblock {\em The theory of matrices. {V}ol. 1}.
\newblock AMS Chelsea Publishing, Providence, RI, 1998.
\newblock Translated from the Russian by K. A. Hirsch, Reprint of the 1959
  translation.

\bibitem{Heath}
D.~Heath, S.~Resnick, and G.~Samorodnitsky.
\newblock {Heavy tails and long-range dependence in on/off processes and
  associated fluid models}.
\newblock {\em Mathematics of Operations Research}, 23:145--165, 1998.

\bibitem{Sigman99}
T.~Huang and K.~Sigman.
\newblock teady-state asymptotics for tandem, split-match and other feedforward
  queues with heavy-tailed service.
\newblock {\em Queueing Systems}, 33:233--259., 1999.

\bibitem{Jelenkovic}
P.~R. Jelenkovic and A.~A. Lazar.
\newblock {Asymptotic results for multiplexing subexponential on-off sources}.
\newblock {\em Advances in Applied Probability}, 31:394--421, 1999.

\bibitem{Kim_09}
J-K. Kim.
\newblock {\em Tail asymptotics of queueing networks with subexponential
  service times}.
\newblock PhD thesis, Georgia Institute of Technology, 2009.

\bibitem{LiMa}
N.~Likhanov and R.~R. Mazumdar.
\newblock Loss asymptotics in large buffers fed by heterogeneous long-tailed
  sources.
\newblock {\em Adv. in Appl. Probab.}, 32(4):1168--1189, 2000.

\bibitem{Nain}
Z.~Liu, P.~Nain, D.~Towsley, and Z-L. Zhang.
\newblock {Asymptotic behavior of a multiplexer fed by long-range dependent
  process}.
\newblock {\em Journal of Applied Probability}, 36(1):105--118, 1999.

\bibitem{Majewski96}
K.~Majewski.
\newblock {\em Large Deviations of Feedforward Queueing Networks}.
\newblock PhD thesis, Ludwig-Maximilian-Universitat, Munchen, 1996.

\bibitem{OCondep97}
N.~O'Connell.
\newblock Large deviations for departures from a shared buffer.
\newblock {\em J. Appl. Probab.}, 34(3):753--766, 1997.

\bibitem{Pakes}
A.~G. Pakes.
\newblock On the tails of waiting-time distributions.
\newblock {\em J. Appl. Probab.}, 12:555--564, 1975.

\bibitem{Makowski}
M.~Parulekar and A.~M. Makowski.
\newblock {Tail probabilities for $M/G/\infty$ input processes}.
\newblock {\em Queueing Systems}, 27:271--296, 1997.

\bibitem{Rolski}
T.~Rolski, S.~Schlegel, and V.~Schmidt.
\newblock {Asymptotics of Palm-stationary buffer content distributions}.
\newblock {\em Advances in Applied Probability}, 31:235--253, 1999.

\bibitem{Zwart_2004}
B.~Zwart, S.~Borst, and M.~Mandjes.
\newblock Exact asymptotics for uid queues fed by multiple heavy-tailed on-off
  sources.
\newblock {\em Annals of Applied Probability}, Vol. 14:903--957, 2004.

\end{thebibliography}
\bibliographystyle{plain}

\appendix
\label{lemmaApp}
\section{Proof of Lemma \ref{lemmascale}}

We will show that $r^{n,J}_m$ and $\rho^{n,J}_m$ form a unique fixed point of a
function for a given set of values of $r_n, \rho_n$.
 Let us now describe this function:
Let $q=2 \sum_{n=1}^N l_n$, $s_1=1,\ s_n = 2 \sum_{k=1}^{n-1} l_k,
\ 1 < n \leq N$ and define
\[
\Omega =\{v \in \rit^q \ : \ v_{s_n + j} \in [0,C_{\pi^n_j}],
n=1,\ldots,N, \ j=1,\ldots, l_n-1
 \}
\]
Here vector $v$ corresponds to the transmission rates and average
loads of the flows at all nodes. Now for $J \in \zit_+^N$, define
the function $T^J: \Omega \rightarrow \Omega$  such that
\begin{eqnarray}
T^J(v)_{s_n+j+1} = \frac{v_{s_n + j}} {\max (
\sum_{k \in \N_{\pi^n_{\bar j}}} J_k v_{s_k + i(k,j,n)} + v_{s_k +
l_k + i(k,j,n)}, C_{\pi^n_{\bar j}}  )},
\\
j=0,\ldots, l_n-2, l_n,
l_n+1, \ldots, 2l_n-2 \nonumber
\end{eqnarray}
 where $i(k,j,n)$ is chosen such that
$\pi^k_i = \pi^n_{\bar j}$ with $\bar j = j \ mod(l_n)$.

The function $T^J$ expresses the input rates to a node in terms of
the output rates of the upstream nodes. Since each input is either
an external flow or the output of another node, we must have the
relation $T^J(v) = v$. Now we need to show that this is indeed the
case and for a fixed $2N$ dimensional vector $w$ given by
 $w_n = v_{s_n}$ and $w_{n+N} = v_{s_n+l_n}$ ,
 i.e., for fixed values of the external input rates, this solution is unique. It is easy to
check that $T^J(\Omega) \subset \Omega$. Let $T^J_w$ be equal to
$T^J$ for a fixed $w$, i.e., $T^J_w = T^J|_{\Omega_w}$ with
$\Omega_w = \{v \in \Omega \ : \ v_{s_n} = w_n, \  v_{s_n+l_n} =
w_{n+N} \}$. Now we will show that $T^J_w$ has a unique fixed
point. From an extension of Banach fixed point theorem~\cite[p.187]{Deimling},
it is enough to show that $T^J_w$ is a condensing
mapping which means that for a given metric $d$ and for $u,v \in
\Omega_w$, $d(T^J_w(u), T^J_w(v)) < d(u,v)$. We  choose the metric
$d$ to be the one corresponding to the $L_2$ norm. To prove that
$T^J_w$ is condensing, it is sufficient to show that the
transformation at each node between input and output rates is a
condensing mapping. Indeed $T^J_w$ can be written as a disjoint
sum of such transformations by choosing the appropriate
permutation of $\{v_i\}$. Hence we will consider a generic mapping
of the form
\[F:D \rightarrow D, \ F(v)_j = \frac{v_j C}{\max (\sum_{i=1}^p J_i v_i, C)}   \] where
$D \subset \rit_+^p$ is a compact set for some $p>0$ and $J \in
\zit_+^p$.
 The compactness condition on the domain $D$ is justified because a flow other than
an external input is output of a node and hence it has a bounded
rate. Define $ \norm{v}  = \sum_{i=1}^p J_i v_i$. Take $u,v \in
D$. If $ \norm{u} < C,  \norm{v} <C$, then $F(u) - F(v) = u-v$. If
$ \norm{u}  < C,  \norm{v}  >C$, take $z = tu + (1-t)v, \ t \in
(0,1)$ such that $\norm{z} = C$. Then,
\[F(u) - F(v) = F(u) - F(z) + F(z) - F(v) = u -z +  F(z) - F(v)\]
Now consider the case $ \norm{u} > C, \norm{v} >C$. Then
$\mathcal{J}_F$, the Jacobian of $F$, is given by
\[\mathcal{J}_F(v) = \frac{C}{\norm{v}^2}A \mbox{ where } A_{ii} = \norm{v}-J_iv_i, \
 A_{ij} = -J_jv_i, \ i \neq j\]
Thus $A = \norm{v}I_p - B$ where $B_{ij} = J_jv_i$ and $I_p$ is
the $p \!\times \! p$ identity matrix. Note that the matrix $B$
has rank ``1" and thus it has one eigenvalue at its trace equal to
$\norm{v}$ with the remaining eigenvalues being $0$. By taking an
invertible matrix $G$ such that $G^{-1}BG$ is Jordan
form~\cite{Gantmacher} of $B$, it follows that $G^{-1}AG$ has one
eigenvalue at $0$ and $p-1$ eigenvalues at $\norm{v}$. Therefore
the largest eigenvalue of $\mathcal{J}_F(v)$ is $C/\norm{v}$ which
is less than 1. This implies that $F$ is a condensing mapping.

As mentioned above, $T^J_w$ can be written as the disjoint sum of
$F$ type transformations. Therefore $T^J_w$ is also a condensing
mapping and has a unique fixed point $v^0(w)$ provided that
$v^0(w)$ satisfies $\sum_{n \in \mathcal{M}_{k}} J_n v^0(w)_{s_n +
n_k} + v^0(w)_{s_n + l_n n_k}> C_k$ for at least one node $k$.
Here $n_k$ is chosen such that $ \pi^n_{n_k}=k$. If this does not
hold, then a fixed point must satisfy $v^0(w)_j = w_n$ for $s_n
\leq j < s_n +l_n$ and $v^0(w)_j = w_{n+N}$ for $s_n +l_n \leq j <
s_{n+1}$. Therefore if  $\sum_{n \in \mathcal{M}_{k}} J_n
v^0(w)_{s_n + n_k} + v^0(w)_{s_n + l_n n_k} < C_k$ holds for every
node $k$, then $v^0(w)$ is a unique fixed point. Thus we have
shown that for every $w$,  there exists a unique $v^0(w)$
satisfying $T_w(v^0(w)) = v^0(w)$. In other words, for a given set
of transmission rates $r_n$ and average loads $\rho_n$ of the
external inputs, the corresponding internal ones are uniquely
defined as in equations (\ref{ratescale}) and (\ref{rhoscale}). Furthermore $v^0(w)$ is also
a continuous function of $w$. Indeed assume that this is not true
and there exists
 $n \rightarrow \infty$, $w_n \rightarrow w$ but
$v^0(w_n) \not \rightarrow v^0(w)$. But since $v^0(w_n)$ lies in a
compact region, there exists $\bar v$ s.t. $v^0(w_{n_k})
\rightarrow \bar v \in \Omega_w$ and $\bar v \in \Omega_w$. Since
$T$ is continuous, $T(\bar v)= \bar v$. But this is in
contradiction to the uniqueness of the fixed point in $\Omega_w$
and thus proves the continuity of $v^0(.)$.
\qed

\end{document}